\documentclass[preprint,eqsecnum,aps,showpacs]{revtex4}
\usepackage{dcolumn}
\usepackage{graphicx}
\usepackage{amsmath}
\usepackage{amssymb}
\usepackage{epsfig}
\usepackage{float}
\usepackage{latexsym}
\usepackage{rotating}

\begin{document}

\title{\small{Primeval acceleration and bounce conditions within induced gravity}}

\vspace{4.0cm}
\author{Nils M. Bezares-Roder$^{1}${\footnote{Nils.Bezares@gmail.com}},
Hemwati Nandan$^{2}${\footnote{hnandan.ctp@jmi.ac.in}} and
Umananda Dev Goswami$^{3}${\footnote{umananda2@gmail.com  \\ \\
{** Dedicated to 75th birth anniversary of Heinz Dehnen.}}}}
\vspace{0.5cm}
\affiliation{\small{$^1$Fachbereich Physik, Universit\"{a}t Konstanz, M677,
78457 Konstanz, Germany\\
$^2$Centre for Theoretical Physics, Jamia Millia Islamia, New Delhi, 110\,025,
India\\
$^3$Department of Physics, Dibrugarh University, Dibrugarh-786\,004, Assam,
India}}

\begin{abstract}
A model of induced gravity with Higgs potential is analysed for the FLRW
cosmology. Conditions of acceleration and signatures for the primeval universe along with the inflation are discussed. It is shown that the scalar-field excitations act quintessentially within effective pressure terms for a negative deceleration parameter. The violation of energy conditions and primeval acceleration appears naturally in the present model in view of the notions for inflationary universe with the avoidance of the Big Bang singularity. \\

\vspace{0.2cm}

\end{abstract}

\vspace{1.0cm}
\maketitle
\section{Introduction}

The models of induced gravity in terms of Jordan--Brans--Dicke (JBD) theories \cite{JBD} lie in the direction of Kaluza--Klein (KK) theories \cite{Kaluza} and theories of everything (TOE) in view of the unified framework of the fundamental interactions in nature \cite{Maeda}. In fact, in scalar--tensor theories (STTs) of induced gravity the scalar field which is coupled to the curvature of spacetime plays the role of an effective gravitational coupling \cite{Maeda}.\\
Incorporating the concept of spontaneous symmetry breaking (SSB) to the
gravity within a STT \cite{Fujii74,Zee79} leads to the breaking of a unified gauge theory into strong, weak and electromagnetic interactions with the scalar field having some vacuum expectation value (VEV). The gravitational coupling in such a broken symmetric theory of gravity is then given in terms of the VEV of a self-interacting scalar field. Furthermore, the potential density of the Bergmann--Wagoner (BW) class of STTs leads to a cosmological term of anti-gravitational nature which lacks in the usual JBD models of gravity.\\
In such models, a potential-density term related to screening of gravity and a local character of the gravitational coupling may be of relevance in the context of Dark Matter, which comprises about 90\% of the whole matter density in the Universe. On the other hand, a cosmological term may be relevant in view of Dark Energy or Quintessence (comprising about 70\% of the whole energy density of the Universe). The nature of both the dark sectors is still unclear.\\
Further, in addition to these problems, Big Bang cosmology leaves some other problems unresolved (\emph{viz} magnetic monopoles, Horizon and Flatness). These problems may be explained without the necessity of an extremely fine tuning of the Universe by means of a highly accelerated expansion (deSitter) era of so-called Inflation \cite{Starobinsky,Guth}, and related to the scalar fields usually called inflatons. The inflationary era is usually characterised as New and Chaotic Inflation depending on the evolution of the scalar field \cite{Albrecht,Linde82,Accetta85,Pollock,Accetta89,Fakir}.\\
The use of a Higgs scalar field in an induced  model of gravity is of quite significant importance due to the role of Higgs potential in mass generation mechanisms of elementary particles \cite{Cervantes95a, Cervantes95b,Dehnen90}. In fact, the Higgs field with its quartic potential density interacts in a gravitational Yukawa form with the states which acquire mass as an immediate consequence of the symmetry breaking \cite{Dehnen90,Dehnen91,Bezares08}, and the model also leads to a time-dependent cosmological-function term of anti-gravitational behaviour (Quintessence). Using the Higgs field, it is possible to build a model of Higgs cosmology \cite{Dehnen92, Dehnen93, Bij94}
in which Inflation may be unified to the mass generation mechanism and hence to the standard model (SM) or a grand unified theory (GUT) in particle physics. Various issues concerning the dark matter phenomenology \cite{Gessner92,Bezares07-DM,Cervantes07-DM,Bezares10-JHEP}, stellar and black hole behaviour \cite{Bezares07-BH,Bezares10-CQG} and inflation
mechanism \cite{Cervantes95a,Cervantes95b,Bezares09-W,Bezares-Dis} have been investigated time and again in detail by using this model.\\
In this article, we investigate the cosmological consequences of an induced gravity model with the Higgs potential in view of the
Friedmann--Lema\^{i}tre--Robertson--Walker (FLRW) spacetime. In
section \ref{model}, we briefly review the model action, field equations and appearance of the cosmological function. In section \ref{flrw}, the
Raychaudhuri and Friedman equations and the deceleration parameter are
derived, followed by a detailed discussion on the quintessential aspects of this model. In section \ref{buniverse}, we also investigate the energy conditions and their violation leading to a bouncing universe scenario in detail.  We also present the evolution of the scale factor in section \ref{numerical} to visualise the inflationary dynamics following the bounce. In the last section, a summary of the important results obtained is discussed in a systematic way.

\section{The Model: Field Equations and Cosmological Function} \label{model}
We consider the following action of scalar--tensor theories
(Bergmann--Wagoner-type) in the natural system of
units \cite{Dehnen92,Dehnen93,Bezares-Dis},
\begin{align}
  S= \int d^4x \sqrt{-g}\left[\frac{\gamma}{16\pi}\phi^\dagger \phi R+ \frac{1}{2}D_\mu \phi^\dagger D^\mu \phi- {\cal V} (\phi)+ {\cal L}_M(\psi,{\cal A}_\mu,\phi)\right].\label{Lagrangian}
\end{align}
Here $\gamma$ is a dimensionless constant, $\phi$ is a scalar or isovectorial field, $R$ is the Ricci scalar and $g$ is the determinant of the metric $g_{\mu \nu}$. Further, we have a Lagrangian contribution corresponding to matter constituted by the fermionic ($\psi$) and massless bosonic (${\cal A}_\mu$) gauge fields. $D_\mu$ and $\partial_\mu$ represent the covariant and the partial derivatives respectively with respect to the coordinate $x^\mu$. Given the gravitational properties of general Higgs fields \cite{Dehnen90,Dehnen91,Bezares-Dis,Bezares09-W}, the potential
density ${\cal V} (\phi)$ in the Lagrangian (\ref{Lagrangian}) has the form given below,
\begin{align}
  {\cal V}(\phi)= \frac{\mu^2}{2}\phi \phi^\dagger+ \frac{\lambda}{4!}(\phi^\dagger \phi)^2+ {\cal \bar{V}}+ {\cal V}_0,\label{Potential}
\end{align}
where $\mu^2<0$ and $\lambda>0$ are real-valued constants and $\bar{\cal V}= 3\mu^4/(2\lambda)$ is a constant term which normalises the potential as per Zee's assumption \cite{Dehnen92,Dehnen93,Bezares07-DM,Bezares07-BH,Bezares10-JHEP}. The presence of $\bar{\cal V}$ alone in addition to the usual terms of the Higgs-type potential in (\ref{Potential}) avoids the presence of a usual cosmological constant in the spontaneously broken mode of symmetry in
the present model with ${\cal V}_0=0$. In absence of ${\cal V}_0$ in the potential (\ref{Potential}), ${\cal \bar{V}}$ can be interpreted as the height of the potential for $V(\phi=0)$. In view of the deWitt's power counting criterion, such a model is also renormalisable, with a coupling of the Higgs field to matter which is of gravitational strength $O(M/M_P)$ \cite{Frommert91,Bij99}. In general, ${\cal V}_0=-3\gamma \mu^2 \Lambda_0 / (4\pi \lambda)$ in (\ref{Potential}) where $\Lambda_0$ is a cosmological constant. In absence of scalar-field excitations from the ground state which appears naturally in this model, ${\cal V}_0$ leads to the usual cosmological constant as of the $\Lambda$CDM  model \cite{Bezares-Dis}. As a result of the spontaneous symmetry breaking, a gravitational coupling is generated in terms of the ground state value of the scalar field given by $v^2=\phi_0\phi_0^\dagger= -6\mu^2/\lambda$, which can further be resolved as $\phi_0vN \, (N^\dagger N=1) $ by using unitary gauge where $N$ is a constant which satisfies $N^\dagger N=1$ \cite{Fujii74,Zee79,Bij94,Bij99}. Further, the scalar-field excitations ($\xi$) are given as $\phi=v(1+ \xi)^{-1/2}N$ \cite{Bezares07-DM}.\\
In order to understand the coupling of the scalar field with the fermionic fields, let us discuss the following form of the Lagrangian
in (\ref{Lagrangian}) used for our investigations (\emph{cf.}
\cite{Bezares10-JHEP}),
\begin{align}
  {\cal L}_M(\psi,\phi)= -\frac{1}{16\pi}{\cal F}_{\mu \nu}{\cal F}^{\mu \nu}+ \frac{i}{2}\,\bar{\psi}\,\gamma^\mu_{_{L,R}}\,D_\mu\psi + h.c.- (1- \hat{q})\,k \,\bar{\psi}_{_R}\,\phi^\dagger \,\hat{x}\,\psi_{_L}+ h.c.,
\end{align}
where ${\cal F}_{\mu \nu}= {\cal A}_{\nu,\mu} - {\cal A}_{\mu,\nu} +
ig [{\cal A}_{\mu}, {\cal A}_{\nu}]$ is the adjoint field-strength tensor where ${\cal A}_{\mu}$ represents the massless bosonic gauge fields in matrix representation and $\psi$ denotes the fermionic field with $L$ and $R$ for left and right-handedness respectively. Here $\hat{x}$ is the Yukawa coupling operator with Yukawa coupling $k$, and $\hat{q}$ represents the coupling of the scalar field to the fermionic fields. The coupling of the scalar field to the fermions has relevant consequences for the source and decay properties of Higgs fields in view of the Klein--Gordon equation for the Higgs field. The parameter $\hat{q}=0$ is related to the Higgs field responsible for the mass generation of elementary particles. However, the Higgs particle here is stable and only interacts gravitationally \cite{Dehnen93,Cervantes95b,Bezares09-W,Bezares-Dis}. The case $\hat{q}=1$ corresponds to a further scalar field acting within astrophysics \cite{Dehnen92,Cervantes95a,Bezares09-W,Bezares-Dis}. Either way, Higgs become essentially stable particles which may possess high length scales related to small masses relevant for flat rotation curves and bar formation (\emph{cf.} \cite{Bij99,Bezares07-BH,Bezares07-DM,Bezares10-JHEP,Bezares10-CQG,Cervantes-RM}).\\
The gravitational strength $\gamma(\gg 1)$ is defined as the square of the ratio of the Planck ($M_P$) and gauge-boson ($M_A$) masses
\cite{Wetterich88,Fujii00}. After symmetry breaking and suppressing the
massless bosonic gauge fields, the scalar-field equation possesses the
following form,
\begin{align}
  D_\mu \partial^\mu \xi+ \frac{\xi}{L^2}= \frac{1}{1+ \frac{4\pi}{3\gamma}}\left[\frac{8\pi G_0}{3}\hat{q} \,T+ \frac{4}{3} \, \Lambda_0\right],\label{scalarfield}
\end{align}
where $T$ is the trace of the energy--stress tensor $T_{\mu \nu}$ related to the matter Lagrangian (${\cal L}_M$) and $L$ is the (Compton) length scale as given below,
\begin{align}
  L= \left[\frac{1+ \frac{4\pi}{3\gamma}}{16\pi G_0 (\mu^4/\lambda)}\right]^{1/2}.
\end{align}
It is basically the inverse of the scalar field mass. Here the gravitational coupling constant $G_0=1/(\gamma v^2)$ is related to a local quantity $\tilde{G}=G_0/(1+\xi)$ induced by field excitations and which appears as effective gravitational coupling (\emph{viz} \cite{Bezares-Dis}).\\
Further, assuming the contribution due to the gauge fields negligible, the energy--stress tensor satisfies the following equation law (\emph{cf.} \cite{Cervantes95a,Cervantes95b,Bezares07-DM,Bezares10-JHEP}),
\begin{align}
  D_\nu T_\mu\,^\nu= (1-\hat{q}) \, \frac{1}{2}\, \partial_\mu \xi (1+\xi)^{-1} \,T.\label{Continuity}
\end{align}
One may immediately notice that for $\hat q=1$, the right hand side of above equation vanishes identically while for $\hat q \neq 1$ the conservation law breaks down by a further \emph {Higgs force term} which is analogue to dynamic extra-dimension force-terms in Kaluza--Klein models (\emph{cf.} \cite{Gu}). On the other hand, Einstein equations read as follows,
\begin{alignat}{1}
  R_{\mu \nu}- \frac{1}{2}&Rg_{\mu \nu}+ \frac{3}{4L^2}\frac{\xi^2}{1+ \xi} g_{\mu \nu}+ \Lambda^*g_{\mu \nu}= -8\pi \tilde{G} T_{\mu \nu}- \frac{1}{1+\xi}\left[D_\nu D_\mu \xi- D_\lambda \partial^\lambda g_{\mu \nu}\right]- \label{Einstein}\\
  &- \frac{\pi}{\gamma}\frac{1}{(1+ \xi)^2}[2\partial_\mu \xi\partial_\nu \xi- \partial_\lambda \xi\partial^\lambda \xi g_{\mu \nu}].\nonumber
\end{alignat}
In equation (\ref{Einstein}), there appears a cosmological-function term which is defined as below,
\begin{align}
  \Lambda^*= 8\pi \tilde{G}{\cal V}(\xi)= \frac{3}{4L^2}\frac{\xi^2}{1+ \xi}+ \frac{\Lambda_0}{1+ \xi}.\label{Cosmological}
\end{align}
This term entails a local contribution which depends on the scalar-field
excitations related to a local character of the gravitational coupling in cosmic evolution and a true cosmological-constant term. For $\Lambda_0\neq 0$ both act anti-gravitationally as related to the quintessential properties or negative pressures or density of a dark sector (\emph{cf.} \cite{Wetterich88,Fujii00}). In order to derive the true cosmological consequences of the present model, we investigate the FLRW cosmology and its consequences of this model in forthcoming sections.

\section{Friedmann Universe and Quintessence} \label{flrw}
Let us consider the Friedmann--Lema\^{i}tre--Robertson--Walker (FLRW) metric,
\begin{align}
ds^2= dt^2- a(t)^2\left[d\chi^2+ f(\chi)^2\left(d\vartheta^2+
\sin^2\vartheta d\varphi^2\right)\right], \label{thp:RW-metric}
\end{align}
\noindent where $\chi$ is the covariant distance and $a(t)$ is the scale
factor. Here $f (\chi) \in \{\sin \chi,\chi,\sinh \chi\}$ is a parameter that depends on spatial curvature $K\in \{1,0,-1\}$ such that the positive, vanishing and negative values of $K$ are attained by three-dimensional spheres, flat space and hyperboloids respectively (\emph{i.e.} the metric (\ref{thp:RW-metric}) essentially corresponds to the three possible spatial geometries). Further, we consider phenomenologically matter as a perfect fluid with the following energy--stress tensor,
\begin{align}
  T_{\mu \nu}= (\epsilon+ p)u_\mu u_\nu- pg_{\mu \nu},
\end{align}
where $\epsilon$ and $p$ represent the energy density and pressure,
respectively, while $u_\mu$ is four-velocity. Since the finite scalar-field excitation in this model imposes a local character of the gravitational coupling, one can write the scalar field excitation $\xi$ and its derivatives in terms of the gravitational coupling as $\xi=G_0/\tilde{G}-1$. In absence of the usual cosmological constant (i.e. $\Lambda_0=0$) and $\gamma\gg 1$, the scalar-field equation
(\ref{scalarfield}) in terms of the effective gravitational coupling now
reads as follows,
\begin{alignat}{1}
  \frac{1}{\tilde{G}^2}\left(\ddot{G}\tilde{G}- 2\dot{G}^2\right)+ 3\frac{\dot{a}}{a}\frac{\dot{G}}{\tilde{G}}+ \frac{1}{L^2}\left(\frac{\tilde{G}}{G_0}- 1\right)=-\frac{8\pi \tilde{G}}{3}\hat{q}(\epsilon- 3p),\label{sceq}
\end{alignat}
while the cosmological function (\ref{Cosmological}) in terms of the effective gravitational coupling is given by
\begin{align}
   \Lambda(\tilde{G})= \frac{3}{4L^2} \left(-1+\frac{\tilde{G}}{G_0}\right)\frac{\tilde{G}}{G_0},
\end{align}
which possesses only non-negative values. However, the continuity
equation (\ref{Continuity}) may now be written as below,
\begin{align}
    \dot{\epsilon}+ 3\frac{\dot{a}}{a}\left(\epsilon+
    p\right)=-(1-\hat{q})\frac{1}{2}\frac{\dot{G}(\xi)}{G(\xi)}
    \left(\epsilon- 3p\right).\label{thp:Kontinuität}
\end{align}
For $\hat{q}=0$, small time deviations of the effective coupling means a
small-valued source within the continuity condition. There appear
entropy-production processes which, however, become minimal when the effective gravitational coupling tends to constant behaviour. For scalar fields with tendency to a constant term, entropy production vanishes.\\
The generalised Einstein equations (\ref{Einstein}) with
(\ref{thp:RW-metric}) now lead to the Friedmann and Raychaudhuri equations respectively as follows,
\begin{alignat}{1}
    \frac{\dot{a}^2+ K}{a^2}=& \frac{8\pi\tilde{G}}{3}\epsilon_T\label{thp:fm1}\\
    \frac{\ddot{a}}{a}=& -\frac{4\pi \tilde{G}}{3}\left(\epsilon_T+ 3p_T\right).\label{thp:fm2}
\end{alignat}
In view of the usual form of these equations in GR, we have defined a total pressure and energy-density distribution as given below,
\begin{align}
  p_T= p+ p_\Lambda ; \,\, \, \, \, \, \quad \epsilon_T= \epsilon+ \epsilon_\Lambda,
\end{align}
which comprises both the usual density and scalar-field contributions to
density ($\epsilon_\Lambda$) and pressure ($p_\Lambda$) as follows,
\begin{alignat}{1}
   \epsilon_\Lambda =\,& {\cal V}+ \frac{3H}{8\pi G(\xi)}\frac{\dot{G}(\xi)}{G(\xi)},\,\label{thp:rlambdag}\\
    p_\Lambda =\,& -{\cal V}- \frac{1}{8\pi G(\xi)}\left[\frac{\ddot{G}(\xi)}{G(\xi)}+ 2\left(H\frac{\dot{G}(\xi)}{G(\xi)}- \frac{\dot{G}(\xi)^2}{G(\xi)^2}\right)\right].\label{thp:plambdag}
\end{alignat}
Within the above-mentioned scalar-field equation of state (EOS) parameters we find the cosmological function which is related to the potential density ${\cal V}(\xi)\equiv {\cal V}$. It is also possible to define a pressure term $p_G=\ddot{\xi}/2$ in the Raychaudhuri equation  (\ref{thp:fm2}) such that
\begin{align}
  \frac{\ddot{a}}{a}= -4\tilde{G}\left(\frac{1}{3}\epsilon+ p+ p_G\right)+ \frac{\Lambda}{3},\label{appag}
\end{align}
where $\epsilon_T+ p_T=\epsilon+ 3p+ 3p_G+ \Lambda$, and with $p_G$ equal to the scalar-field components without potential terms.\\
Dividing equation (\ref{thp:fm2}) by equation (\ref{thp:fm1}) leads to an effective deceleration parameter of the present model which comprises a
curvature term as given below,
\begin{align}
    \tilde{q}= \frac{\ddot{a}a}{\dot{a}^2+ K}= -\frac{1}{2} \, (1+ 3 w_T),\label{thp:aapa}
\end{align}
with the total EOS parameter $w_T=p_T/\epsilon_T$ which includes additional terms due to the scalar-field contribution. The proper deceleration parameter is defined as usual by $q=\ddot{a}/(aH^2)=\tilde{q}(1+ K/\dot{a}^2)$.\\
Usual matter yields a deceleration parameter $q\geq 1$. Thus, any expanding universe should have a decreasing Hubble parameter whereas the local expansion of space is decelerated. This is the case for $w\geq -1/3$. However, super novae observations \cite{Riess98} indicate that the Universe is accelerating (Quintessence) and hence $q$ is negative. The
deceleration parameter (\ref{thp:aapa}) is negative for scalar-field
dominance. It can be easily seen that without the usual matter ($\epsilon=0$), $w_T$ becomes negative (see \cite{Bezares-Dis}). Hence, $p_\Lambda$ is quintessential so that $w_T<w$. The latter may be used for an analysis of the current Universe in terms of cosmological parameters. In the present work, however, we will focus on quintessential, hence anti-gravitational properties of the primeval Universe.

\section{Violation of energy conditions and Bounce} \label{buniverse}

\subsection{Energy Conditions}
\noindent It is reasonable to expect that the energy--stress tensor would satisfy certain conditions such as positivity of the energy density and
dominance of the energy density over pressure. Such requirements are embodied in the following energy conditions \cite{Poisson02,Wald},
\begin{alignat}{2}
  &\text{Weak: }\quad \quad \quad T_{\alpha \beta}v^\alpha v^\beta \geq 0,\,\quad \quad \text{\emph{i.e.}}&\quad \epsilon \geq 0,\, \epsilon+ p_i>0&\label{wec}\\
  &\text{Null: }\quad \quad \quad \,\, T_{\alpha \beta}k^\alpha k^\beta\geq 0,\,\quad \quad \text{\emph{i.e.}}& \epsilon+ p_i\geq 0&\\
  &\text{Strong: }\quad \quad \quad (T_{\alpha \beta}- \frac{1}{2}Tg_{\alpha \beta})v^\alpha v^\beta \geq 0,\,\text{\emph{i.e.}}\,& \epsilon+ \sum_i p_i \geq 0,\quad \epsilon+ p_i\geq 0& \label{sec}\\
  &\text{Dominant: }\quad -T^\alpha\,_\beta v^\beta\, \text{future directed,}  \,\text{\emph{i.e.}}& \epsilon\geq 0,\quad \epsilon\geq |p_i|& \label{dec}
\end{alignat}
where $v^\mu$ is an arbitrary timelike vector which represents the
four-velocity of an arbitrary observer in spacetime, and $k^\mu$ is an
arbitrary, future-directed null vector.\\
The weak energy condition states that the energy density of any matter
distribution as measured by any observer in spacetime must be non-negative. The null energy condition states the same as the weak condition, however with a future-directed null vector $k^\alpha$ instead of $v^\alpha$. The weak energy condition in fact advocates that for every future-pointing timelike vector field, density of matter observed by the corresponding observer would always be positive or non-negative. The strong energy condition does not imply the weak energy condition, yet it appears to be a stronger physical requirement to assume the condition (\ref{sec}) instead of (\ref{wec}) \cite{Poisson02,Wald}. Further, another energy condition (the dominant energy condition) (\ref{dec})
is believed to hold for physically reasonable matter. Under the validity of the dominant energy condition, the mentioned quantity would be a
future-directed vector in addition to the weak energy condition \cite{Wald}.\\
Energy conditions are typically valid for classical matter. They may, however, be violated by quantised matter fields such as within the Casimir effect. Hence, such effect could be used to produce a locally negative-mass region of spacetime as related to a form of effective exotic matter \cite{Morris88}. However, within GR with usual non-exotic matter, the energy conditions are valid in cosmology. Yet, in view of (\ref{thp:aapa}) and $w_T<w$, this may change if
we consider the scalar fields into account. In this way, there are terms
$w_i<0$ which contribute as Dark Energy to density or, in primeval dynamics, to Inflation within a deSitter epoch. For energy conditions to break, however, quintessential contributions have to dominate the dynamics. We analyse the validity of the energy conditions (\ref{wec})--(\ref{dec}) for the present model in the next sections.

\subsection{Inflationary Universe and Bounce}
Negative (effective) pressures or pressure-like terms may lead to
anti-gravitational behaviour such as Quintessence, which is usually identified with a scalar field. Such fields may lead to Inflation as a highly accelerated-expansion era parting from a static condition of the Universe \cite{Starobinsky,Guth,Albrecht,Linde82,Cervantes95a,Cervantes95b}.
The Big Bang singularity scenario is closely associated to the issues related to the energy conditions, type of inflation and initial value of the scalar-field excitation. The scalar-field EOS terms as given by
equations (\ref{thp:rlambdag}) and (\ref{thp:plambdag}) are highly effective. Thus, scalar-field contributions may have rather strong implications on energy conditions and dynamics of the primeval universe.\\
The strong and weak energy conditions together are often referred to as
Penrose--Hawking conditions, and in view of their validity there would be no accelerations $\ddot{a}>0$ \cite{Penrose65}. Further, given the
concaveness of $a(t)$ for all times under the validity of these conditions, $a(t)$ must be equal to zero at some time in the past (\emph{i.e.} there appears a singularity, namely the Big Bang!). In all homogeneous and isotropic models for which the Penrose--Hawking conditions are valid ($-1/3\geq w\geq 1$), a Big Bang singularity is unavoidable. However, it is noteworthy that Yukawa interactions of the magnitude of the nuclear density appear in primeval dynamics \cite{Dehnen75} which might lead to negative pressures having an important role in early stages of the Universe. In usual dynamics of GR, the Big Bang may be avoided in models with exotic matter with dominant negative pressure with $p\geq -\epsilon/3$ as concaveness of $a(t)$ is then no longer valid throughout. These terms in fact act anti-gravitationally. Within induced gravity, anti-gravitative properties may appear especially for a dominance of the scalar field ($p_\Lambda < 0$). If such terms appear and dominate at early stages of the Universe, they may lead to the breaking of Penrose--Hawking energy conditions. An interaction analogous to the Yukawa interaction in \cite{Dehnen75} would be related to $p_\Lambda$, which is related to the potential density ${\cal V}(\xi)$ and the evolution of the gravitational coupling with the Yukawa-interacting Higgs field. In order to check any such violation of energy conditions, let us take a general time $t=t_q$ with the following properties,
\begin{align}
    a(t_q)\neq 0,\quad \dot{a}(t_q)=0;\quad \epsilon(t_q)=0.\label{condsin}
\end{align}
Here $\dot{a}(t_q)=0$ is a condition for inflation at $t=t_q$ for which the Hubble parameter vanishes. $t_q$ shall be identified with $t\approx 0$.\\
The Friedmann equation (\ref{thp:fm1}) then yields
\begin{align}
    \frac{K}{a^2(t_q)}= \frac{1}{4L^2}\frac{G(\xi(t_q))}{G_0}\left(\frac{G_0}{G(\xi(t_q))}- 1\right)^2.\label{kca2}
\end{align}
With the given properties of $\xi$ and $a$, the Universe has to be closed or flat \emph{i.e.} $K\geq 0$. However, $K=0$ is only valid if both
$\dot{\xi}(t_q)=0$ and $\xi(t_q)=0$ are valid ($\dot{G}$ and $\tilde{G}=G_0$ with $\Lambda=0$ for $\Lambda_0=0$). There is $K=1$ (in line with WMAP 5-year results with $\Omega_T=1.099^{+0.100}_{-0.085}$ \emph{cf.} \cite{WMAP5}). For brevity, let us write
\begin{align}
  G(\xi(t_q))\equiv \tilde{G}_q \equiv G_q\quad \text{and}\quad a(t_q)\equiv a_q
\end{align}
and equivalently all other quantities. The Friedmann equation (\ref{kca2}) for $K=1$ now leads to
\begin{align}
  \frac{\tilde{G}_q}{G_0}\left(\frac{\tilde{G}_q}{G_0}-1\right)= 4 \frac{L^2}{a_q^2}. \label{thp:x2a}
\end{align}
The scalar-field excitations for $t=t_q$ now reads as follows,
\begin{align}
    \frac{G_0}{\tilde{G}_q}= -1+ \frac{2L^2}{3}\Lambda_q\left(1\pm \sqrt{1+ 3\frac{\Lambda_q^{-1}}{L^2}}\right),\label{x0}
\end{align}
where $\Lambda_q={3}/{a_q^2}$. Let us focus on the positive signature in
(\ref{x0}). The continuity condition (\ref{thp:Kontinuität}) for $t=t_q$ can be re-written as given below,
\begin{align}
    \dot{\epsilon}_q=-(1-\hat{q})\,\frac{3}{2}\, \frac{\dot{G}_q}{\tilde{G}_q}p_q,\label{dr0}
\end{align}
where $\dot{x}_q\equiv \dot{x}_{|t=t_q}$ is used throughout in our formalism.\\
$\dot{\epsilon}_q$ is to be zero for sign changing to be given at $t=t_q$. For $\hat{q}=1$, this cannot be forced as a condition, as it is directly given. For $\hat{q}=0$, there must be either $\dot{G}_q=0$ or $p(t_q)=0$ for minimal density at $t=t_q$ in order to ensure $\dot{\epsilon}_q=0$ (see section \ref{statsf}).\\
The scalar-field equation (\ref{sceq}) for $t=t_q$ leads to
\begin{align}
    \ddot{\xi}_q= -\left[\frac{\ddot{G}_q}{\tilde{G}_q}- 2\left(\frac{\dot{G}_q}{\tilde{G}_q}\right)^2\right] \frac{G_0}{\tilde{G}_q}= -\frac{1}{L^2}\left(-1+ \frac{G_0}{\tilde{G}_q}\right)- \kappa_0\hat{q}p_q,\label{xpp}
\end{align}
where $\kappa_0=8\pi G_0$. With equation (\ref{x0}) the equation (\ref{xpp}) now has the following simplified structure,
\begin{align}
    \ddot{\xi}_q=- \frac{2}{a^2_q}\left[1+ \sqrt{1+ \frac{a^2_q}{L^2}}\right]-\hat{q}\kappa_0p_q. \label{xpp1}
\end{align}
It is evident from equation (\ref{xpp1}) that for $\hat{q}=0$ (or $p(t_q)=0$), $\ddot{\xi}(t_q)$ is necessarily negative.\\
The Raychaudhuri equation (\ref{thp:fm2}) with $K=1$ now leads to
\begin{align} 
  \ddot{a}_qa_q=1+ \frac{2}{\left(-1+ G_0/\tilde{G}_q\right)^2}\left(-1+ \frac{G_0}{\tilde{G}_q}- L^2(1- \hat{q})\kappa_0p_q\right),\label{termsacc}
\end{align}
which in terms of the length scales ($a$ and $L$) is analogous to the following, 
\begin{align}
  \ddot{a}_q a_q= 1+ \frac{a^2_q}{L^2}\left(1+ \sqrt{1+ \frac{a_q^2}{L^2}}\right)^{-1}- \frac{a_q^4}{L^2}\frac{1}{2}(1- \hat{q})\kappa_0 p_q\left(1+ \sqrt{1+\frac{a_q^2}{L^2}}\right)^{-2}.
\end{align}
One may then easily notice different accelerating ($K$ and positive $\xi_q$) and decelerating (gravitationally attractive) terms ($p_q>0$ for $\hat{q}=0$). For $\ddot{a}_q>0$, from equation (\ref{thp:x2a}) the following condition would hold for acceleration,
\begin{align}
  \frac{1}{2}(1- \hat{q})p_q< \frac{\Lambda_q}{3\kappa_0}\left(1+ \sqrt{1+ \frac{a^2_q}{L^2}}\right)+ \frac{L^2}{3\kappa_0}\Lambda_q^2\left(1+ \sqrt{1+ \frac{a^2_q}{L^2}}\right)^2.\label{accelc}
\end{align}
Acceleration is given unless there are very high pressures $p_q$ in case of $\hat{q}=0$. For $\hat{q}=1$ acceleration appears naturally \emph{i.e.} independent of the pressure terms. Geometrically, this means that (\ref{condsin}) does not give an inflexion point but an extremum, and that this extremum represents a minimum of scale of the universe (a Bounce!). A maximum (decelerating phase) with small length scale $a_q$ is only a physical option when directly followed by a new minimum before the universe collapses in itself. In view of (\ref{accelc}), further, $t=t_q$ leads to a violation of the Penrose--Hawking condition for
\begin{align}
  (1-\hat{q}) p_q<\frac{2\Lambda_q}{3\kappa_0}\left(1+ \sqrt{1+ \frac{a^2_q}{L^2}}\right)=\frac{\Lambda_q}{3\kappa_0} \left(\frac{G_0}{\tilde{G}_q}-1\right).\label{accelPH}
\end{align}
The cosmological function together with the second order derivative of the scalar-field excitation/gravitational coupling (given as a pressure term $p_G$, \emph{cf.} equation (\ref{appag})) act as negative pressure terms or density of exotic matter. For $\hat{q}=1$ they lead to a breaking of the Penrose--Hawking condition independently of $p(t_q)$. For $\hat{q}=0$ relatively high values of the initial pressure are necessary for deceleration to appear (\emph{i.e.} a maximum in cosmic dynamics). Further, for negative total pressures there is an accelerating phase of the universe at $t=t_q$. It is evident from equations (\ref{accelc}) and (\ref{accelPH}) that there is a cosmic acceleration and a breaking of the Penrose--Hawking condition at $t=t_q$ for $\hat{q}=1$. However, for $\hat{q}=0$, acceleration always leads to a breaking of the energy conditions. Furthermore, for large length scales (\emph{i.e.} $L$) in relation to the scale factor $a(t_q)$, both appear more naturally. A singularity is therefore not necessary and a bounce is possible.\\
A bouncing universe has been an object of research for quite some time, often referred to within the context of a ``Phoenix'', oscillatory or cyclic universe and going back to the Lema\^{i}tre \cite{Lemaitre33} universe. The first semi-analytical solution with the filling of a massive scalar field appears in \cite{Starobinsky-Bounce}. Currently, several models and solutions for a bouncing geometry, also regarding symmetry breaking of gauge symmetry of conformal scalar fields, within GR as well as higher dimensional and ekpyrotic-type brane theories have been proposed and analysed \cite{Melnikov79,Khoury01,Steinhardt02, Alimi05,Frampton06,Penrose06,Baum07,Falciano08,Dzhunshaliev09}.
Furthermore, bouncing cosmologies also appear within the recent
Ho\v{r}ava-Lifshitz models of cosmology \cite{Brandenberger09,Czuchry10,Gao10,Maeda10} and Loop Quantum Gravity (LQG) \cite{Date05,Bojowald07,Ashtekar08,Mielczarek08,Helling09,Ashtekar10} such that a (Gamow) Bounce joins contracting (pre-big-bang) and expanding (post-big-bang) cosmological branches by a (quantum) bridge.

\subsection{The maximal scalar-field and energy conditions}\label{statsf}
\noindent In order to analyse the initial conditions of density, we need to derive the initial properties of the scalar field. In equation (\ref{dr0}), we have the energy condition for which density has to possess an extremal value in order to acquire sign changing at $t=t_q$. Let us now consider the time derivative of the Friedmann equation,
\begin{align}
  2\frac{\dot{a}}{a}\left(\frac{\ddot{a}}{a}- \frac{\dot{a}^2}{a^2}- \frac{K}{a^2}\right)=\left[\frac{8\pi \tilde{G}}{3}\left(\dot{\varrho}+ \varrho\frac{\dot{G}}{\tilde{G}}\right)- \frac{\dot{a}}{a}\left(\frac{\dot{a}}{a}\frac{\dot{G}}{\tilde{G}}- \frac{\ddot{G}}{\tilde{G}}+ \frac{\dot{G}^2}{\tilde{G}^2}\right)- \frac{\ddot{a}}{a}\frac{\dot{G}}{\tilde{G}}\right]+ \dot{\Lambda}.\label{tdfe}
\end{align}
For $t=t_q$ the equation (\ref{tdfe}) leads to
\begin{align}
  0=\frac{\ddot{a}(t_q)}{a(t_q)}\frac{\dot{G}_q}{\tilde{G}_q}+ \dot{\Lambda}(t_q).  \label{appxl}
\end{align}
Using (\ref{termsacc}), the derivative of the cosmological function from
equation (\ref{appxl}) now reads as
\begin{align}
  \dot{\Lambda}(t_q)= \frac{1}{a^2_q}\left[1+ \frac{2}{(-1+G_0/\tilde{G}_q)^2}\left((-1+G_0/\tilde{G}_q)- L^2(1- \hat{q})\kappa_0p_q\right)\right]\frac{\dot{G}_q}{\tilde{G}_q}.\label{accelerrr}
\end{align}
According to equation (\ref{accelerrr}), acceleration is related to
$\dot{\Lambda}_q$ when $\dot{\xi}_q\neq 0$. Further, $\dot{G}_q>0$ leads
to $\dot{\Lambda}_q>0$. On the other hand, the usual definition of $\Lambda$ leads to
\begin{align}
  \dot{\Lambda}=-\frac{3}{4L^2}\frac{\dot{G}}{\tilde{G}_q}\left(-1+ \frac{G_0}{\tilde{G}}\right)\left(1+ \frac{\tilde{G}_q}{G_0}\right).\label{accelerrr1}
\end{align}
For $t=t_q$, $\dot{G}_q>0$ means $\dot{\Lambda}_q<0$. Hence, for the
consistency of equation (\ref{accelerrr}) with (\ref{accelerrr1}) at $t=t_q$ we need
\begin{align}
  \dot{G}_q \,\equiv\, 0.\label{g00}
\end{align}
Thus, one obtains automatically $\dot{\epsilon}_q\equiv 0$ regardless the initial pressure and fermionic coupling $\hat{q}$. Now, from the equations (\ref{thp:rlambdag}) and (\ref{thp:plambdag}) for $\xi_q\gg 1$ and $a_q\ll L$,
\begin{alignat}{1}
  \epsilon_\Lambda=& {\cal V}_q= \frac{\Lambda_q}{\kappa_0}>0,\\
  p_\Lambda=& -\frac{5 \Lambda_0}{3\kappa_0}<0,\\
  \epsilon_\Lambda+ p_\Lambda=& -\frac{2}{3}\frac{\Lambda_q}{\kappa_0}<0.
\end{alignat}
For $\epsilon_q=0=p_q$, Inflation conditions lead to a rupture of all the energy conditions as described in equations (\ref{wec})--(\ref{dec}). Further, according to the equations (\ref{xpp}), (\ref{xpp1}) and (\ref{g00}), the gravitational coupling grows at $t=t_q$ as the scalar-field excitation $\xi$ falls from its maximal value. Hence, for $t=t_q$ the scalar field leads to Chaotic Inflation
with the scalar-field excitations falling to the ground state. This is in consistency with the rupture of the energy conditions and the initial Bounce.

\subsection{The Planck-length Bounce}\label{PLB}
\noindent At this point of consideration it is still open that which value $t_q\approx 0$ acquires exactly. We may assume $a_q\ll L$ especially with a length scale $L$ which possesses a value of galactic range \cite{Bezares07-DM,Cervantes-RM,Bezares10-JHEP}. Let us assume that the contraction of the Universe for higher redshifts goes on until the following uncertainty relation for energy is valid,
\begin{align}
    \Delta E \Delta t = \hbar.
\end{align}
At this scale, quantum mechanics becomes dominant, and time itself is not exactly determined anymore as classical mechanics lose its validity. Hence, let us further assume that this point gives an initial condition such that we consider the Planck time $t_P\approx 0$. At this time, quantum fluctuations persist on the scale of the Planck length $L_P = c\, t_{P}$ that may be regarded as related to a minimal scale of the Universe. Therefore, there is $t_q=t_P$. Further, in consequence we
have $a(t_P)\equiv a_P\cong L_P$, which represents the order of magnitude of the cosmological horizon at $t=t_P$, which is $L_P\sim 10^{-33}$cm.\\
Following the Friedmann and Raychaudhuri equations, the Planck density
$\varrho_{_P}$ is of the order $(G_0t_P^2)^{-1}$. Given that energy density $\epsilon(t_q)$ is assumed as vanishing, the Planck mass would be constituted by pressure terms $p(t_P)$ and scalar-field excitations $\xi(t_P)\equiv \xi_P$. Consequently, for matter, we have
\begin{align}
    \varrho(t_P)\neq \varrho_P.
\end{align}
With $a(t_P)=L_P$, the scalar field excitations ($\xi_P$) are expressed as
\begin{align}
   \xi(t_P)\equiv \xi_P\equiv -1+ \frac{G_0}{\tilde{G}_P}\cong \frac{2L^2c^3}{G_0\hbar}\left(1+ \sqrt{1+ \frac{G_0\hbar}{L^2c^3}}\right).\label{xiattp}
\end{align}
For $\xi_P\gg 1$ ($\tilde{G}\rightarrow 0$), the equation (\ref{xiattp}) leads to
\begin{align}
    \xi_P\cong \frac{4L^2c^3}{G_0\hbar}\cong \frac{4L^2}{L_p^2} \approx 10^{66}\text{cm}^{-2}\cdot L^2\,.\label{thp:xi0pg}
\end{align}
For a scale factor $L$ of galactic range, the scalar-field excitation acquires very large values. Further, the cosmological term $\Lambda(\xi_P)\equiv \Lambda_P$ reads
\begin{align}
  \Lambda_P=& \frac{3}{4L^2}\frac{\xi_P^2}{1+ \xi_P}\approx 10^{66}\text{cm}^{-2}.\label{thp:Lp}
\end{align}
The same result is achieved directly from the Friedmann equation (\ref{thp:fm1}) with $\Lambda_P/3= K/L_P^2$ and $\epsilon(t_P)=0$.
However, we obtain an effective density of the system solely with $a(t_P)=L_P$ which is hence related to the Planck mass\index{Planck scale}. Using equation (\ref{thp:Lp}) with the Friedmann equation (\ref{thp:fm1}), the Planck density reads as given below,
\begin{align}
  \varrho_P=\frac{3\,c^2}{8\pi G_0L_P^2}\cong 10^{93}\frac{\text{g}}{\text{cm}^3}\,.
\end{align}
The Planck density does indeed appears as effective density at $t_P$ although $\varrho(t_P)=0$. Thus, the Friedmann equation is consistent with an initial density $\varrho(t_P)$ to be vanishing for $\dot{a}_P=0$. The Planck density and hence the Planck mass are given by the scalar field at $t=t_P$, or more exactly by the scalar-field potential at the Planck time, given by the Planck length itself. The Friedmann equation (\ref{thp:fm1}) now reads as
\begin{align}
  \frac{1}{L_P^2}=\frac{8\pi}{3} \frac{G_0}{c^4}\varrho_P=\frac{\Lambda_P}{3}\,. \label{thp:fm1e}
\end{align}
Using (\ref{thp:fm1e}), the Raychaudhuri equation by using the
expressions for $\ddot{\xi}_P$ and $\xi_P$ with $K=1$ is now given as below,
\begin{align}
  2\frac{\ddot{a}_P}{L_P}- \left(2+ \frac{L_P^2}{L^2}\right)\frac{c^2}{L_P^2}= -2\frac{L_P^2}{L^2}\frac{\pi G_0}{c^2}(1- \hat{q})p(t_P)\,.\label{aplp}
\end{align}
For vanishing values of the initial pressure $p(t_P)$ or $\hat{q}=1$, the right-hand side of equation (\ref{aplp}) disappears and cosmic acceleration at $t=t_P$ is inevitable. For $L_P\ll L$, the first term on the right hand side of equation (\ref{aplp}) is dominant and we have
\begin{align}
  \ddot{a}_P\sim 10^{53}\text{cm s}^{-2},\quad (\text{for $L_P \ll L$})\,.
\end{align}
However, for $L\cong 10^{22}$cm, $L_Pc^2/L^2\cong 10^{-57}$cm s$^{-2}$. As in the general case $t=t_q$ for $\hat{q}=0$, positive pressures pull acceleration down since the pressure acts gravitationally. The pressure term needed for restoration of energy conditions and deceleration, however, is dependent on the reciprocate value of the squared length scale $L$ and on the squared value of the Planck length. Yet, even for length scales of the order of magnitude of the Planck length, the pressure needed for deceleration and prohibiting a bounce is extremely high. Further, it seems more natural that $p$ be zero at
$t=t_q$ given that $\varrho(t_P)$ is zero. Thus, acceleration is given at this initial state of the Universe.\\
For length scales relevantly larger than the Planck length, the equation (\label{aplp}) leads to
\begin{align}
  \ddot{a}_P \approx \frac{c^2}{L_P}\,,
\end{align}
which is independent of $\hat{q}$, $L$ and $p(t_P)$. Further, the second derivative of the field excitations reads as
\begin{align}
  \ddot{\xi}_P=-2\frac{c^2}{L_P^2},\label{ddx}
\end{align}
which is also a leading counter-gravitational factor. For $L\gg L_P$, it is clear that the energy conditions are therefore broken. This is related to a highly accelerated state for the primeval Universe at $t\approx 0$ with a Big Bounce. There is a very high cosmological function $\Lambda_P$ which acts anti-gravitationally and leads to the accelerated expansion. After the bounce, following the equations (\ref{g00}) and (\ref{ddx}), the scalar-field excitations and the cosmological function fall towards its ground-state value while the scale factor grows.\\
Within the mechanism of Inflation, a cosmic evolution with $\phi\ll v\rightarrow v$ is related to New Inflation \cite{Cervantes95a,
Cervantes95b} while $\phi\gg v\rightarrow v$ is related to Chaotic Inflation. Given the high excitations $\xi$ for $t=t_P$, it may be assumed that Chaotic Inflation appears within this analysis. Further, given that there is no input from a pre Big-Bang region, the appearing inflation is a case of Super-Inflation \cite{Ashtekar10}. Such situation is in agreement with the shown breaking of all energy conditions and the initial highly accelerated state at $t=t_P$. In order to have a comprehensive understanding of the above-mentioned facets of the
present model, we make a numerical study  of the evolution of the scale factor
in the next section.

\section{Evolution of Scale factor} \label{numerical}
It is important to investigate the evolution of the scale factor for the scale $t>t_P$ in order to know if there appears any kind of rollover contraction (even though $t_P$ itself shows acceleration) or if acceleration persists. The investigations on the evolution of scale factor are also necessary to mark the presence of any inflationary epoch within this model. In view of unification and of the stable nature of Higgs particles as in the present model, we consider the afore-mentioned values of the length scale for our analysis of the scale factor \cite{Bezares07-DM,Cervantes-RM,Bezares10-JHEP}. According to earlier analyses \cite{Gessner92,Bezares07-DM}, a length scale of the order of magnitude of galactic bulges (in order of kpc) may lead to flat rotation curves. Further, for the strongest bar formation in isolated and interacting galaxies \cite{Cervantes-RM}, a similar value is
derived  with $L$ of the order 10kpc (\emph{i.e.} a mass of $M\approx
10^{-26}$eV/$c^2$). Finally, the scalar-field contributions of density
dominate within a dark-matter profile for $L\approx a_{flat}/36$ with
$a_{flat}$ as the distance at which rotation curves become flat
\cite{Bezares10-JHEP}. Hence, we consider  $L=10^{22}$cm. We further consider $a_q=10^{-33}$cm$ \;\approx a_P$ (further lengths are intended especially for comparison) and the results from section \ref{PLB} as initial conditions for the fields. In order to visualise the inflationary epoch followed by Bounce, we prevent the evolution of the cosmic acceleration, cosmic velocity and scale factor ratio respectively for the case $\hat q=1$.\\
\begin{figure}[h] \centering
    \includegraphics[width = 7.5cm, height = 7.5cm]{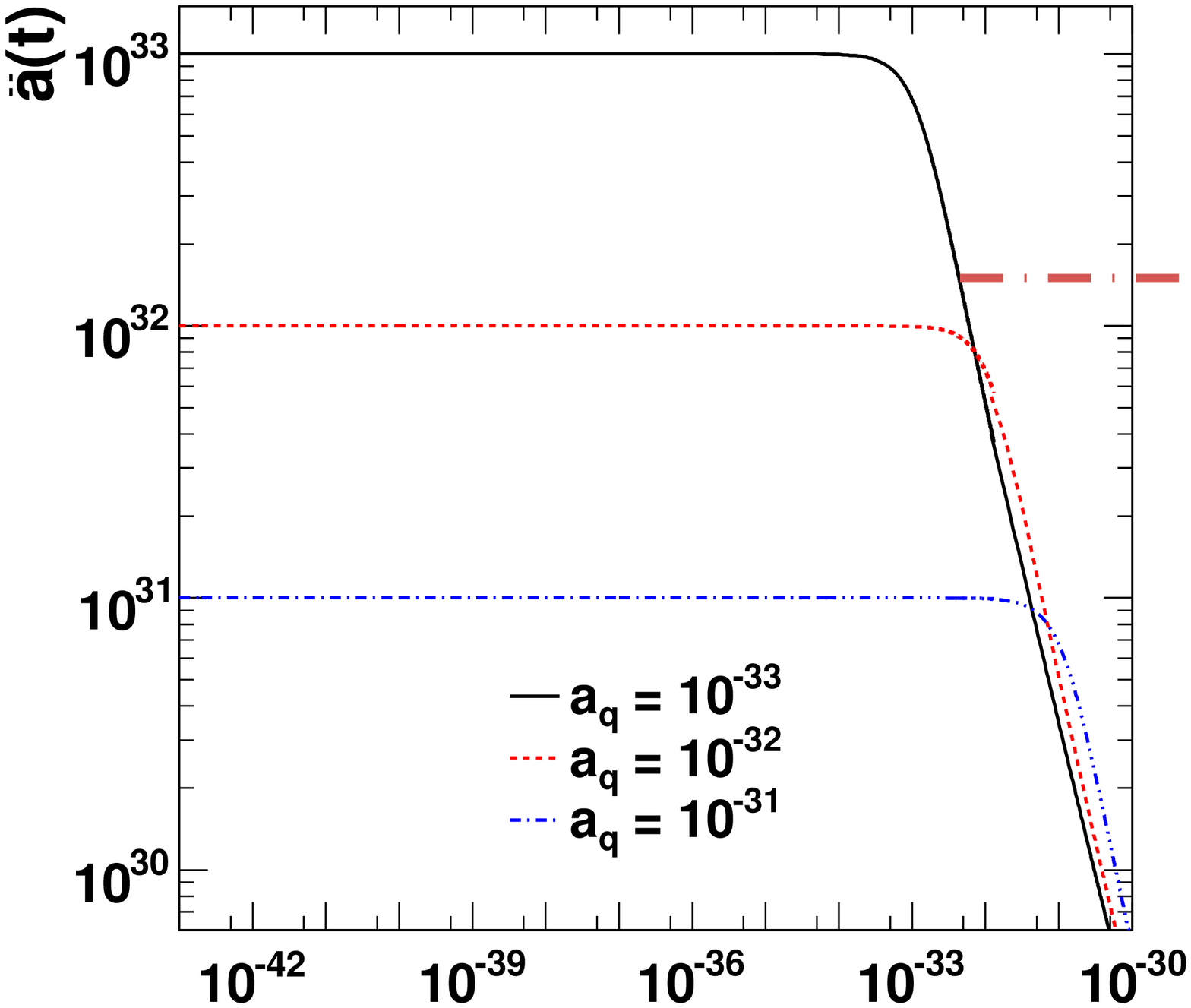}
    \hspace{-0.33cm}
    \includegraphics[width = 7.5cm, height = 7.5cm]{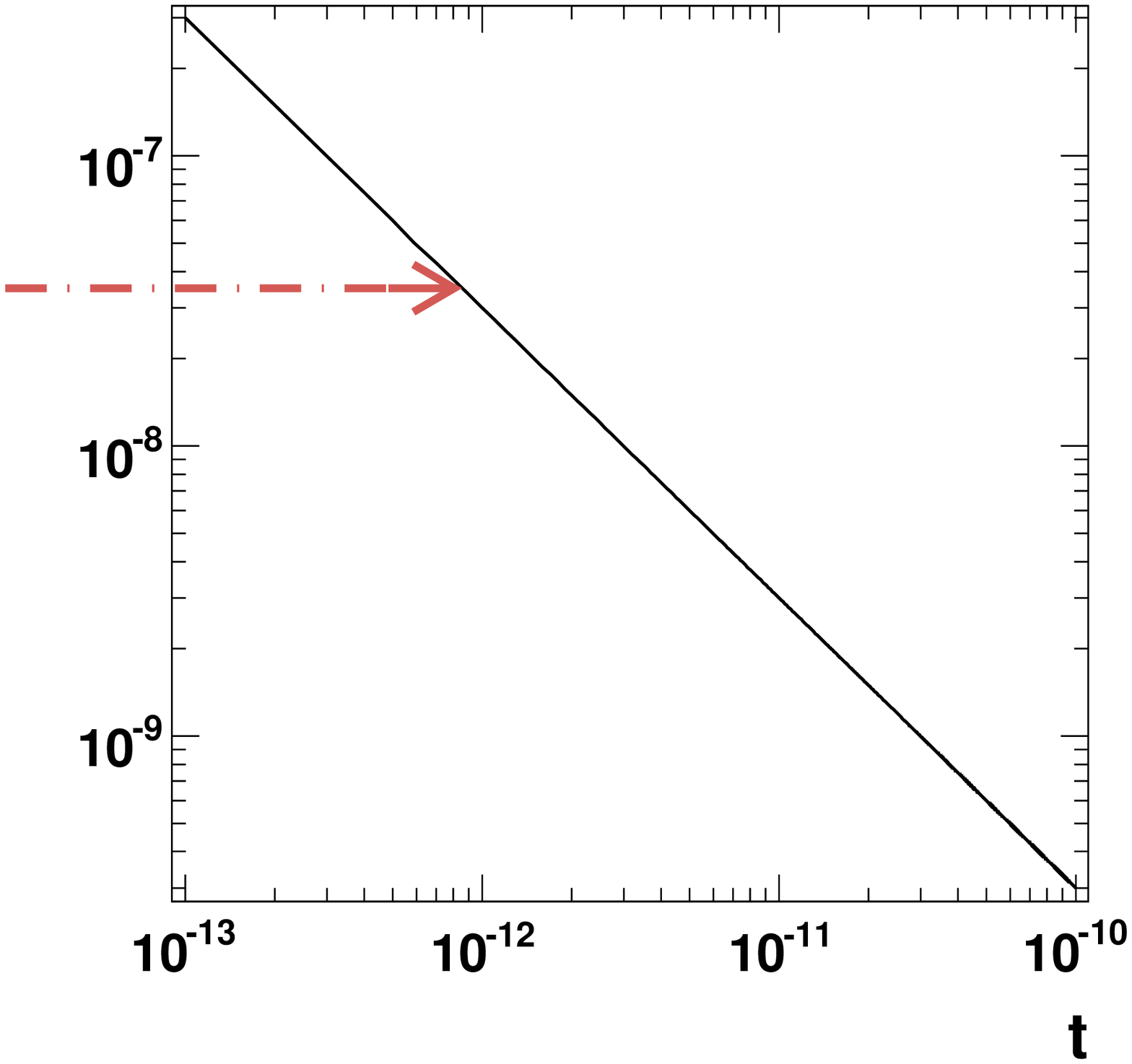}
    \caption{Primeval cosmic acceleration (in cm/s$^2$) after the Bounce
(time in seconds). The plots are obtained for the length scale $L = 10^{22}$cm and $\hat{q}$ = 1. The right-hand panel of the figure shows the later time behaviour of cosmic acceleration after Bounce. Same conditions are also applicable to the cases of Figure \ref{vel} and \ref{scale} for corresponding situations.}
    \label{accel}
    \end{figure}
The cosmic acceleration as presented in the Figure \ref{accel} clearly shows a highly accelerated phase of the universe after the Bounce with no upcoming rollover contraction for the scale $t>t_P$. It also indicates a constant behaviour which is higher for lower initial values of $a(t)$. At a time near $t \approx 10^{-33}$sec the acceleration starts falling.
\begin{figure}[h] \centering
    \includegraphics[width = 7.5cm, height = 7.5cm]{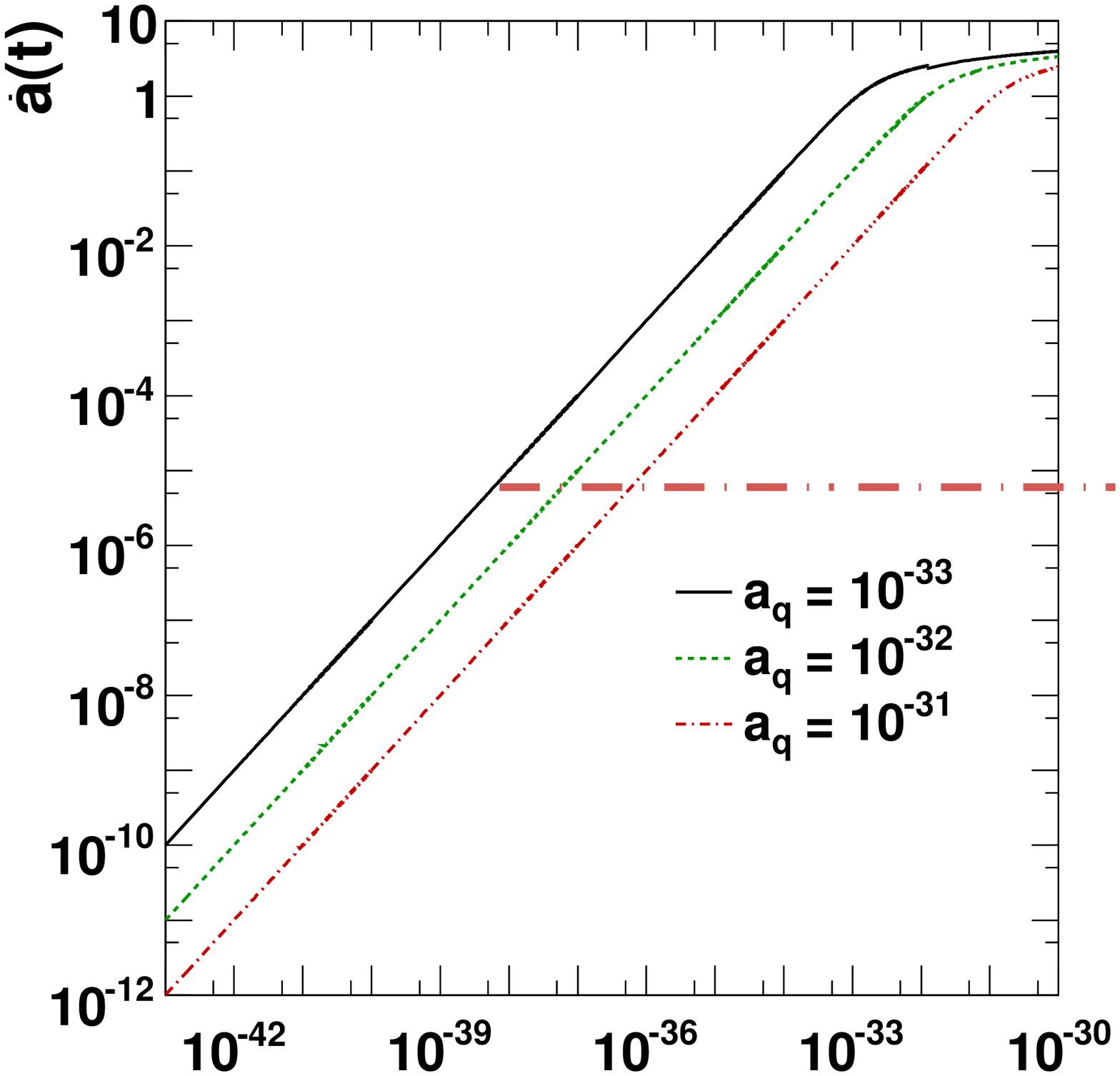}
    \hspace{-0.33cm}
    \includegraphics[width = 7.5cm, height = 7.5cm]{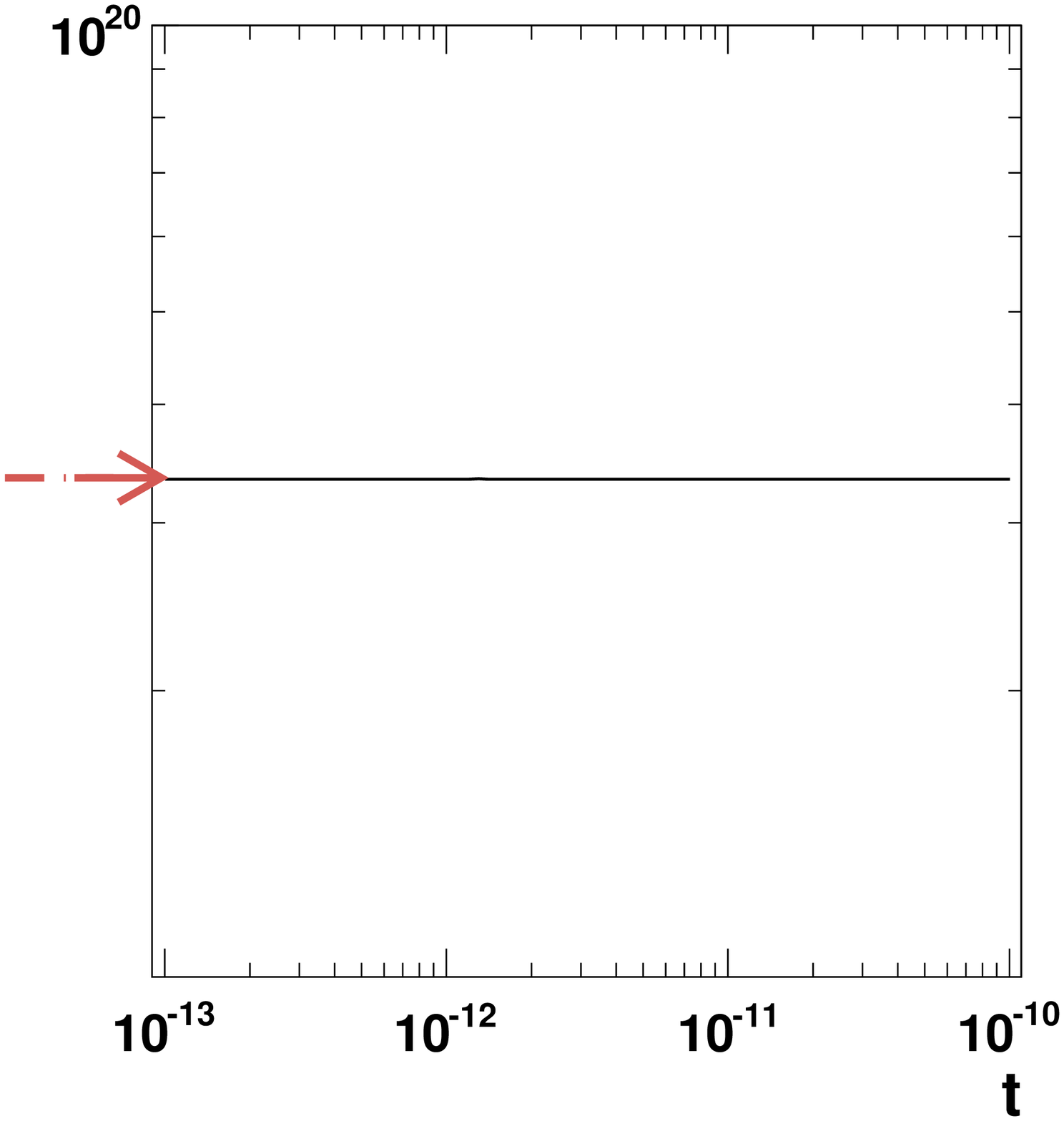}
    \caption{Primeval cosmic velocity (in cm/s) after the Bounce.}
    \label{vel}
    \end{figure}
The same can be seen in the Figure \ref{vel} in form of the slope of the
cosmic velocity which is related to a primeval Hubble rate. The cosmic
velocity grows rapidly up to
$t\lesssim 10^{-33}$sec. Cosmic acceleration falls for a cosmic velocity over a value of unity, while for 
$t\gtrsim 10^{-33}$sec there remains a (flatter) exponential growth such that high cosmic velocities are visible at 
later times (see the right panel of the Figure \ref{vel}).
   \begin{figure}[h] \centering
    \includegraphics[width = 7.5cm, height = 7.5cm]{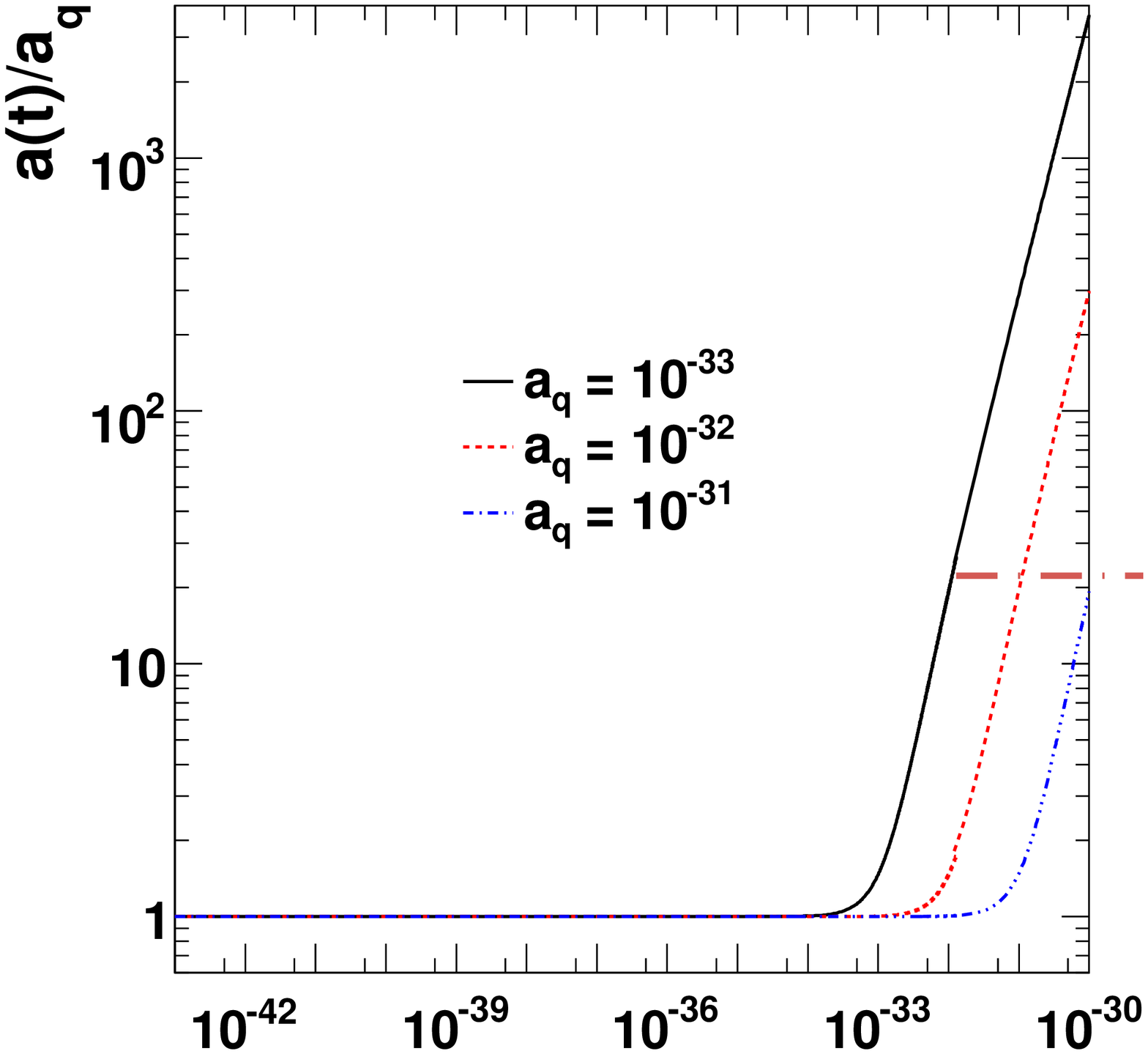}
    \hspace{-0.33cm}
    \includegraphics[width = 7.5cm, height = 7.5cm]{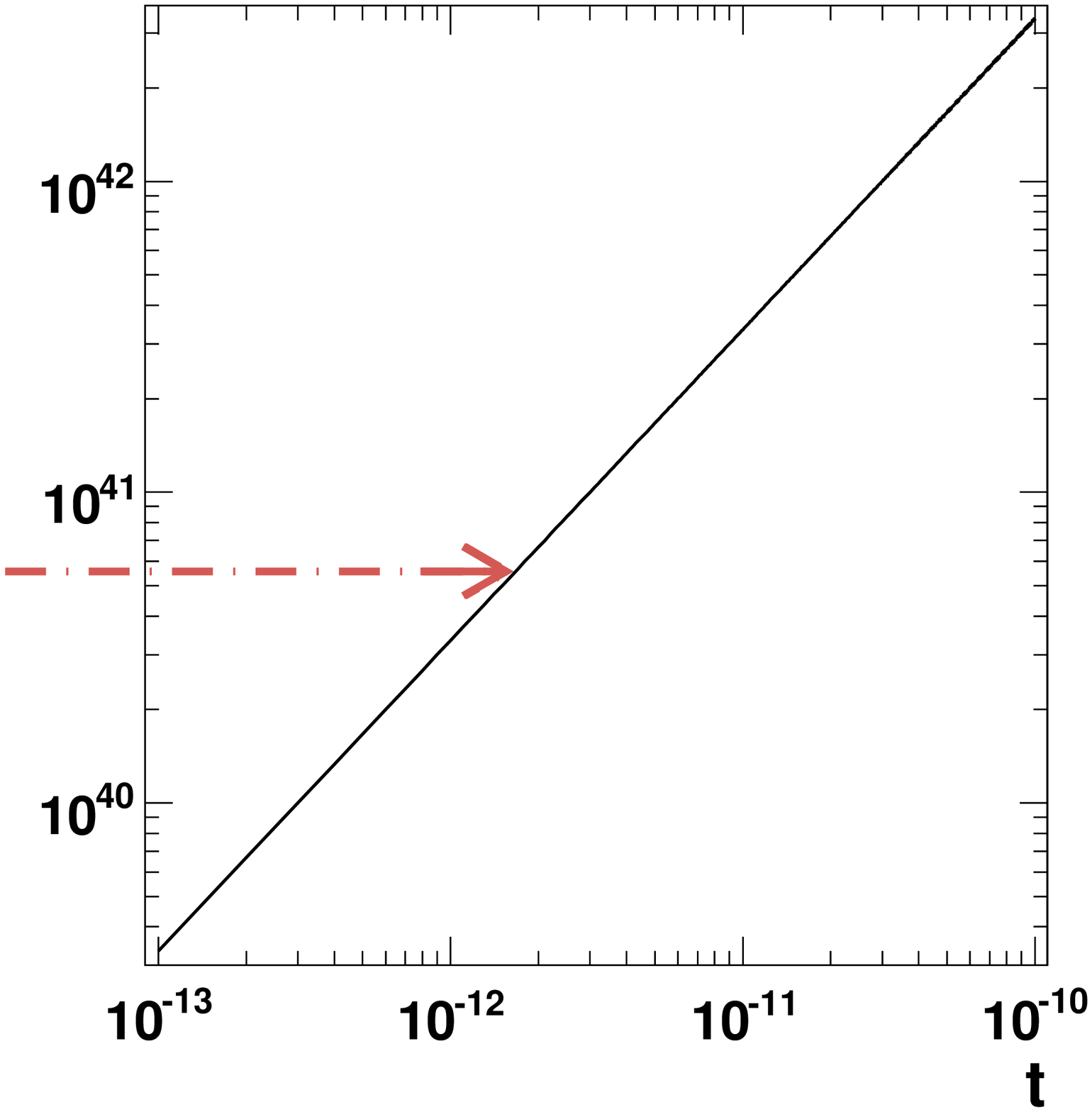}
    \caption{Evolution of the scale-factor ratio after the Bounce.}
    \label{scale}
    \end{figure}
The ratio of the scale parameter (\emph{i.e.} $a(t)$) to the initial scale (\emph{i.e.} $a_q$), as visible in the Figure \ref{scale}, indicates a relatively weak growth up to 
$t_{infl}\approx 10^{-33}$sec. The scale $t_{infl}$ may then be interpreted as the time where Inflation actually starts. However,
up to a scale $t_e > 10^{-13}$sec 
the e-folds of Inflation ($a_{t_e}/a_{t_{infl}}$) yield over 40. Though, the exact number of such e-folds (\emph{i.e.} higher than 40 to completely suppress spatial curvature in the present day's observable Universe) is still an open issue in view of the considered values of the different parameters for the present analysis. Further, a detailed investigation of the scalar field evolution for the case $\hat q=0$ (\emph{i.e.} when the fermions are coupled to the Higgs field) is still needed.

\section{Summary and Conclusions} \label{summary}
In this article, we have investigated Bounce and Inflation conditions for a model of induced gravity with Higgs potential. We have especially investigated the role of Higgs field excitations and an the effective gravitational coupling in deriving the conditions for Bounce followed by Inflation. We consider both the scenarios of mass-generating Higgs fields and of further, cosmological Higgs fields in view of FLRW cosmology. To conclude, we provide below a concise summary of the results obtained.

\begin{itemize}
  \item The total energy and the pressure are given by usual matter and scalar-field terms entailing the potential and dynamical terms of the gravitational coupling. The scalar field indeed acts as a negative pressure for Quintessence and for the scalar-field dominance, there appears a negative deceleration parameter.
\end{itemize}
\begin{itemize}
  \item Inflation condition leads to a closed Universe with $K=1$ for vanishing (minimal) energy density and finite scale factor as initial conditions which naturally include a maximal value of the scalar field. For these initial conditions, the acceleration appears naturally with negative effective pressures. Furthermore, the energy conditions are broken and a Bounce occurs without the appearance of a Big Bang singularity.
  \item The decaying scalar-field excitation after Bounce leads to a high effective gravitational coupling together with diminishing cosmological-function terms. The effective quintessential pressure terms are given by a decaying cosmological function and further gravitational-coupling terms which do not vanish for vanishing cosmological-function terms.
\end{itemize}
\begin{itemize}
  \item The Planck density is given by the effective density of the scalar field. High scalar-field excitations and acceleration appear together with a large cosmological constant related to the reciprocal value of the squared Planck length as scale factor. Such excitations may be given within Chaotic Inflation, which refers to higher initial values of the scalar fields.
  \item Evolution after the scale $t_P$ indicates that there is no rollover contraction, but rather an ongoing highly accelerated state which after some time leads to a rapid, exponential growth of the scale factor of the universe with over 40 e-folds within a time period of $10^{-10}$ seconds of Inflation.
\end{itemize}
The exact conditions for which there appears a contracting phase of the oscillatory universe, leading to a Bounce then followed by an inflationary phase in the present model are still needed to investigate in greater detail especially by constraining the parameters of the Higgs potential in view of the observations.\\
Finally, the essential goal behind this work has been to demonstrate some cosmological consequences of the early Universe derived from a gravitational theory coupled to the isovectorial Higgs field in the standard model and grand unified theories in particle physics. Thus, we have an interesting approach to have a convergence between the Higgs field of particle physics and facets of the primeval universe from the viewpoint of cosmology.

\section*{Acknowledgments}
\small{The authors are grateful to Prof. H. Dehnen for his persistent
intellectual support, constant encouragement and guidance for the
scientific research time and again. One of the authors (HN) is thankful to the University Grants Commission, New Delhi, India for financial support under the UGC--Dr. D. S. Kothari post-doctoral fellowship programme (Grant No. F-2/2006(BSR)/13-187/2008(BSR). HN is also thankful to Prof. M. Sami for his kind motivation.}

\end{document}